\documentclass[longauth]{aa}

\usepackage{euclid}
\usepackage{graphicx}
\usepackage{natbib}
\usepackage{scalerel}

\usepackage[table]{xcolor}

\usepackage{txfonts}
\usepackage[pdfencoding=auto,psdextra]{hyperref}
\hypersetup{
    colorlinks=true,
    linkcolor=blue,
    filecolor=magenta,
    urlcolor=blue,
    citecolor=blue
}
\usepackage{amsmath}	
\usepackage{amssymb}	
\usepackage{xfrac}
\usepackage{mathtools}
\usepackage[utf8]{inputenc}
\usepackage[T1]{fontenc}
\usepackage[normalem]{ulem}

\makeatletter
\renewcommand*\aa@pageof{, page \thepage{} of \pageref*{LastPage}}
\makeatother

\usepackage[utf8]{inputenc}

\usepackage[switch, modulo]{lineno}

\begin{document}
%
%
\title{\Euclid: Calibrating photometric redshifts with spectroscopic cross-correlations\thanks{This paper is published on behalf of the Euclid Consortium.}}

\newcommand{\orcid}[1]{} 
\author{K.~Naidoo\orcid{0000-0002-9182-1802}$^{1,2}$\thanks{\email{krishna.naidoo.11@ucl.ac.uk}}, H.~Johnston\orcid{0000-0002-4074-0308}$^{3,2}$, B.~Joachimi$^{2}$, J.L.~van den Busch\orcid{0000-0001-9059-2553}$^{4}$, H.~Hildebrandt\orcid{0000-0002-9814-3338}$^{5}$, O.~Ilbert$^{6,7,8}$, O.~Lahav$^{2}$, N.~Aghanim$^{9}$, B.~Altieri\orcid{0000-0003-3936-0284}$^{10}$, A.~Amara$^{11}$
% ...
}

\institute{$^{1}$ Center for Theoretical Physics, Polish Academy of Sciences, al. Lotnik\'{o}w 32/46, 02-668 Warsaw, Poland\\
	$^{2}$ Department of Physics and Astronomy, University College London, Gower Street, London WC1E 6BT, UK\\
	$^{3}$ Institute for Theoretical Physics, Utrecht University, Princetonplein 5, 3584 CE Utrecht, The Netherlands\\
	$^{4}$ Ruhr-Universit\"at Bochum, Astronomisches Institut, German Centre for Cosmological Lensing (GCCL), Universit\"atsstr. 150, 44801 Bochum, Germany\\
	$^{5}$ Ruhr University Bochum, Faculty of Physics and Astronomy, Astronomical Institute (AIRUB), German Centre for Cosmological Lensing (GCCL), 44780 Bochum, Germany\\
	$^{6}$ Aix-Marseille Univ, CNRS, CNES, LAM, Marseille, France\\
	$^{7}$ California institute of Technology, 1200 E California Blvd, Pasadena, CA 91125, USA\\
	$^{8}$ Jet Propulsion Laboratory, California Institute of Technology, 4800 Oak Grove Drive, Pasadena, CA, 91109, USA\\
	$^{9}$ Universit\'e Paris-Saclay, CNRS, Institut d'astrophysique spatiale, 91405, Orsay, France\\
	$^{10}$ ESAC/ESA, Camino Bajo del Castillo, s/n., Urb. Villafranca del Castillo, 28692 Villanueva de la Ca\~nada, Madrid, Spain\\
	$^{11}$ Institute of Cosmology and Gravitation, University of Portsmouth, Portsmouth PO1 3FX, UK\\
	$^{12}$ Dipartimento di Fisica e Astronomia, Universit\'a di Bologna, Via Gobetti 93/2, I-40129 Bologna, Italy\\
	$^{13}$ INAF-Osservatorio di Astrofisica e Scienza dello Spazio di Bologna, Via Piero Gobetti 93/3, I-40129 Bologna, Italy\\
	$^{14}$ INFN-Sezione di Bologna, Viale Berti Pichat 6/2, I-40127 Bologna, Italy\\
	$^{15}$ Universit\"ats-Sternwarte M\"unchen, Fakult\"at f\"ur Physik, Ludwig-Maximilians-Universit\"at M\"unchen, Scheinerstrasse 1, 81679 M\"unchen, Germany\\
	$^{16}$ Max Planck Institute for Extraterrestrial Physics, Giessenbachstr. 1, D-85748 Garching, Germany\\
	$^{17}$ Dipartimento di Fisica, Universit\'a degli studi di Genova, and INFN-Sezione di Genova, via Dodecaneso 33, I-16146, Genova, Italy\\
	$^{18}$ INFN-Sezione di Roma Tre, Via della Vasca Navale 84, I-00146, Roma, Italy\\
	$^{19}$ Department of Physics "E. Pancini", University Federico II, Via Cinthia 6, I-80126, Napoli, Italy\\
	$^{20}$ INAF-Osservatorio Astronomico di Capodimonte, Via Moiariello 16, I-80131 Napoli, Italy\\
	$^{21}$ Instituto de Astrof\'isica e Ci\^encias do Espa\c{c}o, Universidade do Porto, CAUP, Rua das Estrelas, PT4150-762 Porto, Portugal\\
	$^{22}$ Dipartimento di Fisica, Universit\'a degli Studi di Torino, Via P. Giuria 1, I-10125 Torino, Italy\\
	$^{23}$ INFN-Sezione di Torino, Via P. Giuria 1, I-10125 Torino, Italy\\
	$^{24}$ INAF-Osservatorio Astrofisico di Torino, Via Osservatorio 20, I-10025 Pino Torinese (TO), Italy\\
	$^{25}$ INAF-IASF Milano, Via Alfonso Corti 12, I-20133 Milano, Italy\\
	$^{26}$ Institut de F\'{i}sica d'Altes Energies (IFAE), The Barcelona Institute of Science and Technology, Campus UAB, 08193 Bellaterra (Barcelona), Spain\\
	$^{27}$ Port d'Informaci\'{o} Cient\'{i}fica, Campus UAB, C. Albareda s/n, 08193 Bellaterra (Barcelona), Spain\\
	$^{28}$ Institut d'Estudis Espacials de Catalunya (IEEC), Carrer Gran Capit\'a 2-4, 08034 Barcelona, Spain\\
	$^{29}$ Institute of Space Sciences (ICE, CSIC), Campus UAB, Carrer de Can Magrans, s/n, 08193 Barcelona, Spain\\
	$^{30}$ INAF-Osservatorio Astronomico di Roma, Via Frascati 33, I-00078 Monteporzio Catone, Italy\\
	$^{31}$ INFN section of Naples, Via Cinthia 6, I-80126, Napoli, Italy\\
	$^{32}$ Dipartimento di Fisica e Astronomia "Augusto Righi" - Alma Mater Studiorum Universit\'a di Bologna, Viale Berti Pichat 6/2, I-40127 Bologna, Italy\\
	$^{33}$ INAF-Osservatorio Astrofisico di Arcetri, Largo E. Fermi 5, I-50125, Firenze, Italy\\
	$^{34}$ Institut national de physique nucl\'eaire et de physique des particules, 3 rue Michel-Ange, 75794 Paris C\'edex 16, France\\
	$^{35}$ Centre National d'Etudes Spatiales, Toulouse, France\\
	$^{36}$ Institute for Astronomy, University of Edinburgh, Royal Observatory, Blackford Hill, Edinburgh EH9 3HJ, UK\\
	$^{37}$ Jodrell Bank Centre for Astrophysics, Department of Physics and Astronomy, University of Manchester, Oxford Road, Manchester M13 9PL, UK\\
	$^{38}$ European Space Agency/ESRIN, Largo Galileo Galilei 1, 00044 Frascati, Roma, Italy\\
	$^{39}$ Univ Lyon, Univ Claude Bernard Lyon 1, CNRS/IN2P3, IP2I Lyon, UMR 5822, F-69622, Villeurbanne, France\\
	$^{40}$ Institute of Physics, Laboratory of Astrophysics, Ecole Polytechnique F\'{e}d\'{e}rale de Lausanne (EPFL), Observatoire de Sauverny, 1290 Versoix, Switzerland\\
	$^{41}$ Mullard Space Science Laboratory, University College London, Holmbury St Mary, Dorking, Surrey RH5 6NT, UK\\
	$^{42}$ Departamento de F\'isica, Faculdade de Ci\^encias, Universidade de Lisboa, Edif\'icio C8, Campo Grande, PT1749-016 Lisboa, Portugal\\
	$^{43}$ Instituto de Astrof\'isica e Ci\^encias do Espa\c{c}o, Faculdade de Ci\^encias, Universidade de Lisboa, Campo Grande, PT-1749-016 Lisboa, Portugal\\
	$^{44}$ Department of Astronomy, University of Geneva, ch. d\'Ecogia 16, CH-1290 Versoix, Switzerland\\
	$^{45}$ INFN-Padova, Via Marzolo 8, I-35131 Padova, Italy\\
	$^{46}$ Universit\'e Paris-Saclay, Universit\'e Paris Cit\'e, CEA, CNRS, Astrophysique, Instrumentation et Mod\'elisation Paris-Saclay, 91191 Gif-sur-Yvette, France\\
	$^{47}$ INAF-Osservatorio Astronomico di Trieste, Via G. B. Tiepolo 11, I-34143 Trieste, Italy\\
	$^{48}$ Aix-Marseille Univ, CNRS/IN2P3, CPPM, Marseille, France\\
	$^{49}$ Istituto Nazionale di Astrofisica (INAF) - Osservatorio di Astrofisica e Scienza dello Spazio (OAS), Via Gobetti 93/3, I-40127 Bologna, Italy\\
	$^{50}$ Istituto Nazionale di Fisica Nucleare, Sezione di Bologna, Via Irnerio 46, I-40126 Bologna, Italy\\
	$^{51}$ INAF-Osservatorio Astronomico di Padova, Via dell'Osservatorio 5, I-35122 Padova, Italy\\
	$^{52}$ Institute of Theoretical Astrophysics, University of Oslo, P.O. Box 1029 Blindern, N-0315 Oslo, Norway\\
	$^{53}$ von Hoerner \& Sulger GmbH, Schlo{\ss}Platz 8, D-68723 Schwetzingen, Germany\\
	$^{54}$ Technical University of Denmark, Elektrovej 327, 2800 Kgs. Lyngby, Denmark\\
	$^{55}$ Max-Planck-Institut f\"ur Astronomie, K\"onigstuhl 17, D-69117 Heidelberg, Germany\\
	$^{56}$ Department of Physics and Helsinki Institute of Physics, Gustaf H\"allstr\"omin katu 2, 00014 University of Helsinki, Finland\\
	$^{57}$ NOVA optical infrared instrumentation group at ASTRON, Oude Hoogeveensedijk 4, 7991PD, Dwingeloo, The Netherlands\\
	$^{58}$ Argelander-Institut f\"ur Astronomie, Universit\"at Bonn, Auf dem H\"ugel 71, 53121 Bonn, Germany\\
	$^{59}$ School of Physics and Astronomy, Faculty of Science, Monash University, Clayton, Victoria 3800, Australia\\
	$^{60}$ Dipartimento di Fisica e Astronomia "Augusto Righi" - Alma Mater Studiorum Universit\`{a} di Bologna, via Piero Gobetti 93/2, I-40129 Bologna, Italy\\
	$^{61}$ Department of Physics, Institute for Computational Cosmology, Durham University, South Road, DH1 3LE, UK\\
	$^{62}$ Universit\'e C\^{o}te d'Azur, Observatoire de la C\^{o}te d'Azur, CNRS, Laboratoire Lagrange, Bd de l'Observatoire, CS 34229, 06304 Nice cedex 4, France\\
	$^{63}$ INFN-Bologna, Via Irnerio 46, I-40126 Bologna, Italy\\
	$^{64}$ European Space Agency/ESTEC, Keplerlaan 1, 2201 AZ Noordwijk, The Netherlands\\
	$^{65}$ Department of Physics and Astronomy, University of Aarhus, Ny Munkegade 120, DK-8000 Aarhus C, Denmark\\
	$^{66}$ Centre for Astrophysics, University of Waterloo, Waterloo, Ontario N2L 3G1, Canada\\
	$^{67}$ Department of Physics and Astronomy, University of Waterloo, Waterloo, Ontario N2L 3G1, Canada\\
	$^{68}$ Perimeter Institute for Theoretical Physics, Waterloo, Ontario N2L 2Y5, Canada\\
	$^{69}$ AIM, CEA, CNRS, Universit\'{e} Paris-Saclay, Universit\'{e} de Paris, F-91191 Gif-sur-Yvette, France\\
	$^{70}$ Space Science Data Center, Italian Space Agency, via del Politecnico snc, 00133 Roma, Italy\\
	$^{71}$ Institute of Space Science, Bucharest, Ro-077125, Romania\\
	$^{72}$ Instituto de Astrof\'isica de Canarias, Calle V\'ia L\'actea s/n, E-38204, San Crist\'obal de La Laguna, Tenerife, Spain\\
	$^{73}$ Departamento de Astrof\'{i}sica, Universidad de La Laguna, E-38206, La Laguna, Tenerife, Spain\\
	$^{74}$ Dipartimento di Fisica e Astronomia "G.Galilei", Universit\'a di Padova, Via Marzolo 8, I-35131 Padova, Italy\\
	$^{75}$  Universit\'e Paris Cit\'e, CNRS, Astroparticule et Cosmologie, F-75013 Paris, France\\
	$^{76}$ Departamento de F\'isica, FCFM, Universidad de Chile, Blanco Encalada 2008, Santiago, Chile\\
	$^{77}$ Centro de Investigaciones Energ\'eticas, Medioambientales y Tecnol\'ogicas (CIEMAT), Avenida Complutense 40, 28040 Madrid, Spain\\
	$^{78}$ Instituto de Astrof\'isica e Ci\^encias do Espa\c{c}o, Faculdade de Ci\^encias, Universidade de Lisboa, Tapada da Ajuda, PT-1349-018 Lisboa, Portugal\\
	$^{79}$ Universidad Polit\'ecnica de Cartagena, Departamento de Electr\'onica y Tecnolog\'ia de Computadoras, 30202 Cartagena, Spain\\
	$^{80}$ Universit\'e de Gen\`eve, D\'epartement de Physique Th\'eorique and Centre for Astroparticle Physics, 24 quai Ernest-Ansermet, CH-1211 Gen\`eve 4, Switzerland\\
	$^{81}$ Kapteyn Astronomical Institute, University of Groningen, PO Box 800, 9700 AV Groningen, The Netherlands\\
	$^{82}$ Infrared Processing and Analysis Center, California Institute of Technology, Pasadena, CA 91125, USA\\
	$^{83}$ IFPU, Institute for Fundamental Physics of the Universe, via Beirut 2, 34151 Trieste, Italy\\
	$^{84}$ INAF-Osservatorio Astronomico di Brera, Via Brera 28, I-20122 Milano, Italy\\
	$^{85}$ Dipartimento di Fisica "Aldo Pontremoli", Universit\'a degli Studi di Milano, Via Celoria 16, I-20133 Milano, Italy\\
	$^{86}$ INFN-Sezione di Milano, Via Celoria 16, I-20133 Milano, Italy\\
	$^{87}$ Institut d'Astrophysique de Paris, 98bis Boulevard Arago, F-75014, Paris, France\\
	$^{88}$ Junia, EPA department, F 59000 Lille, France}

 \date{}

%
%
\abstract{Cosmological constraints from key probes of the \Euclid imaging survey rely critically on the accurate determination of the true redshift distributions, $n(z),$ of tomographic redshift bins. We determine whether the mean redshift, $\langle z \rangle$, of ten \Euclid tomographic redshift bins can be calibrated to the \Euclid target uncertainties of $\sigma(\langle z\rangle)<0.002\,(1+z)$ via cross-correlation, with spectroscopic samples akin to those from the Baryon Oscillation Spectroscopic Survey (BOSS), Dark Energy Spectroscopic Instrument (DESI), and \Euclid's NISP spectroscopic survey. We construct mock \Euclid and spectroscopic galaxy samples from the Flagship simulation and measure small-scale clustering redshifts up to redshift $z<1.8$ with an algorithm that performs well on current galaxy survey data. The clustering measurements are then fitted to two $n(z)$ models: one is the true $n(z)$ with a free mean; the other a Gaussian process modified to be restricted to non-negative values. We show that $\langle z \rangle$ is measured in each tomographic redshift bin to an accuracy of order 0.01 or better. By measuring the clustering redshifts on subsets of the full Flagship area, we construct scaling relations that allow us to extrapolate the method performance to larger sky areas than are currently available in the mock. For the full expected \Euclid, BOSS, and DESI overlap region of approximately $6000\,{\rm deg}^{2}$, the uncertainties attainable by clustering redshifts exceeds the \Euclid requirement by at least a factor of three for both $n(z)$ models considered, although systematic biases limit the accuracy. Clustering redshifts are an extremely effective method for redshift calibration for \Euclid if the sources of systematic biases can be determined and removed, or calibrated out with sufficiently realistic simulations. We outline possible future work, in particular an extension to higher redshifts with quasar reference samples.}

%
%
%
%
\titlerunning{\Euclid: calibrating photometric redshifts with spectroscopic cross-correlations}
\authorrunning{K. Naidoo et al.}

\maketitle
%
%
%
%

\section{Introduction}

The European Space Agency's \Euclid space mission\footnote{\href{https://www.euclid-ec.org/}{https://www.euclid-ec.org/}} \citep{Laureijs2011} will map out the positions of billions of galaxies in the Universe. The imaging survey is designed to measure the flux of galaxies in broadband photometric filters over visual and infrared wavelengths, and spectroscopic redshifts will be determined for a subsample of these galaxies using slitless spectroscopy of emission line galaxies (ELGs). Cosmological parameters and models will be constrained via clustering and weak-lensing measurements, made on galaxy samples split into photometric redshift (photo-$z$) bins. Critical to the accuracy and precision of cosmological parameter constraints is the determination of the true redshift distributions of these tomographic bins \citep[see e.g.][]{Huterer2006, Busch2020}.

A variety of techniques have been developed for this purpose, including the aggregation, or `stacking', of individual galaxy photometric redshift probability distribution functions (PDFs; \citealt{Tanaka2018,Hadzhiyska2020,Ilbert2021}); `direct calibration' of the photometric sample $n(z)$ through the re-weighting of spectroscopic colour-redshift spaces \citep{Lima2008,Hildebrandt2016,Wright2020}; and, the focus of this paper, cross-correlations with spectroscopic samples, or `clustering redshifts' \citep{Newman2008}.

Since the seminal work of \cite{Newman2008}, cross-correlation redshift calibration has seen a fair amount of development \citep{Schmidt2013,Menard2013,McQuinn2013,Sanchez2014,Scottez2016,Morrison2017,Scottez2018,Alarcon2019} but has only recently made its way into state-of-the-art large-scale structure analyses \citep{DESHoyle2018,Gatti2018,GattiDES2021,Hildebrandt2018,DESMyles2021,Cawthon2022}. Clustering redshifts use small angular scales where the signal-to-noise of galaxy clustering is highest, and thus the achievable precision scales directly with the area of photometric--spectroscopic overlap \citep[see][]{Cawthon2018}. Next-generation photometric surveys such as \Euclid and the Rubin Observatory's Legacy Survey of Space and Time (LSST; \citealt{LSST2009}) will be able to employ thousands of square degrees of such overlap, in collaboration with the new generation of spectroscopic surveys, such as the Dark Energy Spectroscopic Instrument \citep[DESI;][]{Aghamousa2016}, allowing for the most powerful application of clustering redshifts to date.

Clustering redshifts are estimated through the angular cross-correlations of galaxy samples. A tracer sample with secure redshift estimates (from spectroscopy or narrow-band photometry) is cross-correlated with a target sample (which typically has only broadband photometry), for which we wish to accurately and precisely determine the redshift distribution, $n(z)$. Underlying the method is the assertion that the on-sky positions of galaxies in the two samples will only be correlated where they inhabit similar ranges in redshift, that is, where their $n(z)$ overlap. Thus, the amplitude of the cross-correlation at a given redshift contains information about the amplitude of the (known) tracer, $n(z),$ and the unknown target, $n(z)$, but is degenerate with the galaxy biases of both the photometric and spectroscopic galaxy samples. For the spectroscopic sample, the bias can be estimated by measuring the auto-correlation function. However, the photometric sample bias is more difficult to constrain and is the primary source of systematic errors for this mode of study \citep{Busch2020}.

Clustering redshift calibration is complementary to the more standard methods of photo-$z$ calibration frequently employed in the literature. Such approaches rely upon real or simulated source photometry, either for characterising colour-redshift relations or for the use of photo-$z$ estimation algorithms. Photometric uncertainties and biases then propagate directly into the $n(z)$ determination. The potential for biased $n(z)$ determination from these techniques can be significant due to degeneracies in the colour-redshift distribution and to the incompleteness of spectroscopic samples in colour-redshift space \citep{Hartley2020,Wright2020}. Clustering redshifts suffer none of these limitations, as they rely instead upon the accuracy and sufficient coverage of point-estimated redshifts. Being susceptible to independent sources of bias, each method is well suited to cross-check the others; \cite{Alarcon2019} combined the photometry- and clustering-based approaches in a hierarchical Bayesian framework, leveraging all available information to recover mean redshifts with uncertainties of $\sim\SI{3e-3}{}$, even in scenarios of poor or biased spectroscopic completeness.

For \Euclid to achieve one of its primary science goals, specifically a figure of merit (FoM) on dark energy of $>400$ \citep{Laureijs2011}, uncertainties on the mean redshift, $\sigma\left(\langle z\rangle \right),$ for each tomographic redshift bin must be less than $0.002\,(1+z)$ \citep[68\% confidence limits;][]{Laureijs2011}. In this paper we assess the potential for photometric-spectroscopic cross-correlations to form a major component of the redshift calibration in \Euclid. We aim to use approximately $402\,{\rm deg}^{2}$ of the \Euclid Flagship simulation \citep{Potter2017}, with realistic photo-$z$ point estimates based upon LSST-like photometry, to forecast the expected uncertainties of clustering redshift calibration as the \Euclid data volume grows. We define realistic spectroscopic samples in the simulation, based upon DESI and Baryon Oscillation Spectroscopic Survey (BOSS) selection criteria \citep{Aghamousa2016,Dawson2013}, and make use of the clustering redshift method \citep{Schmidt2013,Menard2013} as implemented in the \texttt{yet\_another\_wizz (YAW)} software package \citep{Busch2020}. We explore the dependence of $\sigma\left(\langle z\rangle \right)$ upon the area of spectroscopic overlap and extrapolate from $402\,{\rm deg}^{2}$ to the full projected overlap region for \Euclid, BOSS, and DESI. We also explore various methods of mitigation for systematic biases associated with the unknown redshift evolution of the photometric galaxy bias, including the use of spectroscopic auto-correlations and internal consistency checks available only to simulations.

This paper is organised as follows. In Sect. \ref{sec:data} we describe the Flagship simulated data, the realistic photo-$z$ employed in our analysis, and our definitions of spectroscopic sub-samples. Section \ref{sec:method} details our cross-correlation methodology, galaxy bias correction, and the extrapolation of uncertainties to the full \Euclid, BOSS, and DESI overlap. In Sect. \ref{sec:results} we present the results of clustering redshift calibration for these simulations and in Sect. \ref{sec:conclusions} discuss the implications for \Euclid clustering redshifts and avenues to pursue for future research.

\begin{figure*}
	\centering
	\includegraphics[width=17cm]{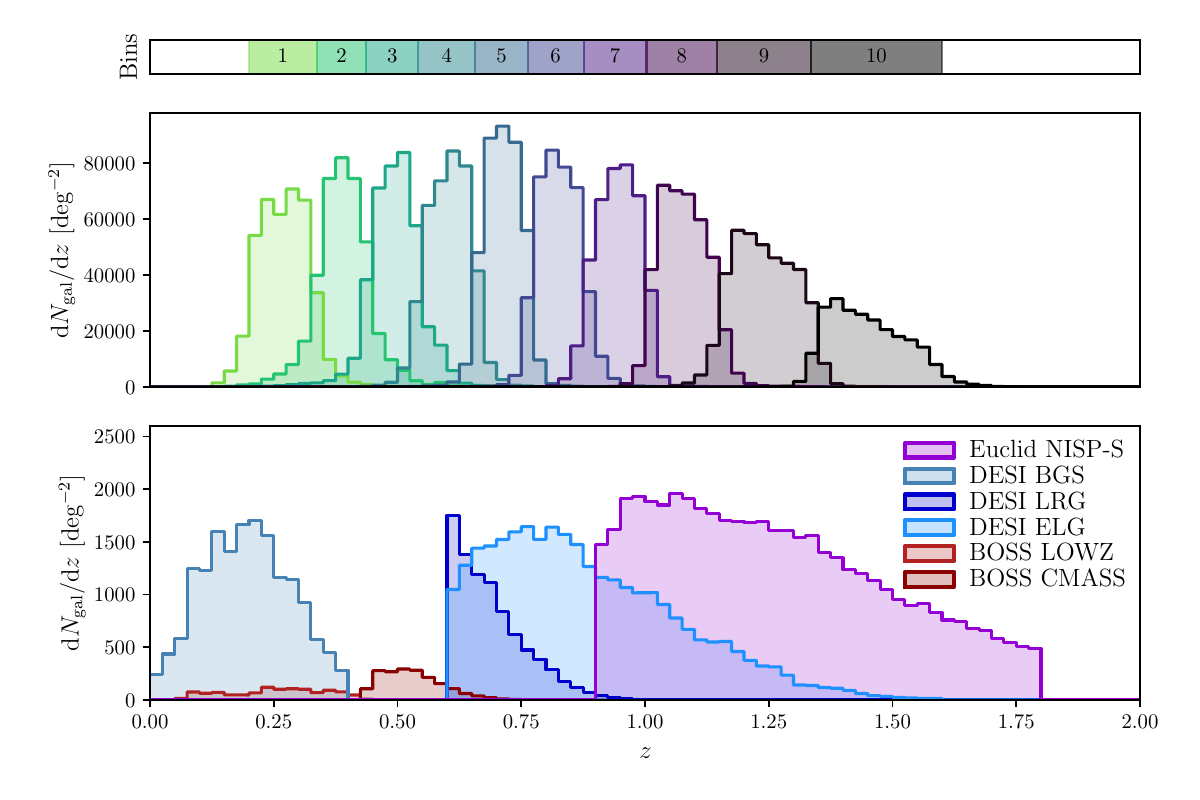}
	\caption{Photometric redshift distribution in comparison to reference spectroscopic samples. \emph{Top}: Redshift ranges for each of the ten tomographic bins. \emph{Middle}: True redshift distribution of each tomographic bin. \emph{Bottom}: Redshift distribution of the reference spectroscopic samples: BOSS-like LOWZ and CMASS; DESI-like BGS, LRG, and ELG; and \textit{Euclid} NISP-S. The reference samples only support redshifts up to $1.8$, and dictate the maximum redshift of bin 10.}
	\label{fig:euclid_photoz_vs_specz}
\end{figure*}

\section{Data: Simulated photometric and spectroscopic samples}
\label{sec:data}

Below we explain our process for constructing simulated \Euclid photometric `target' samples and spectroscopic `tracer' samples used to study clustering redshifts for \Euclid.

\subsection{Summary of photometric and spectroscopic samples}

In this study, we cross-correlate simulated photometric target galaxies with \Euclid-like mock photo-$z$ and mock spectroscopic tracer samples from the BOSS-like LOWZ and CMASS; the DESI-like Bright Galaxy Survey (BGS); luminous red galaxies (LRGs)  and ELGs; and the \Euclid-like NISP-S galaxies. We defined our BOSS- and DESI-like galaxy samples on simulated, noiseless photometry, wherein randomly distributed redshift `failures' are modelled by sparse sampling to appropriate number densities. However, systematic failures arising from noisy photometry, and potentially correlating spatially, or with variable survey depth, were not implemented -- the impact of these should be explored in future work.

In Fig. \ref{fig:euclid_photoz_vs_specz} we display the true redshift distributions for the ten photometric bins, defined using one of the photo-$z$ samples defined in Sec.~\ref{sec_photoz}, and compare them with our mock spectroscopic sample redshift distributions. The lack of spectroscopic samples above redshift $1.8$ sets a hard upper-limit for our inferred target $n(z)$'s. Future studies will benefit greatly from understanding the role that quasar galaxy samples from BOSS, eBOSS \citep{EBOSS2016}, or DESI \citep{DESIQSO2020} can play in better constraining the \Euclid photometric sample at higher redshifts. We also note the relatively small tracer--target redshift overlap for galaxies in the range $0.4$ to $0.6$, where only mock BOSS galaxies are available for cross-correlation; this has the potential to adversely affect the $n(z)$ constraints for this redshift range.

\begin{figure*}
	\centering
	\includegraphics[width=17cm]{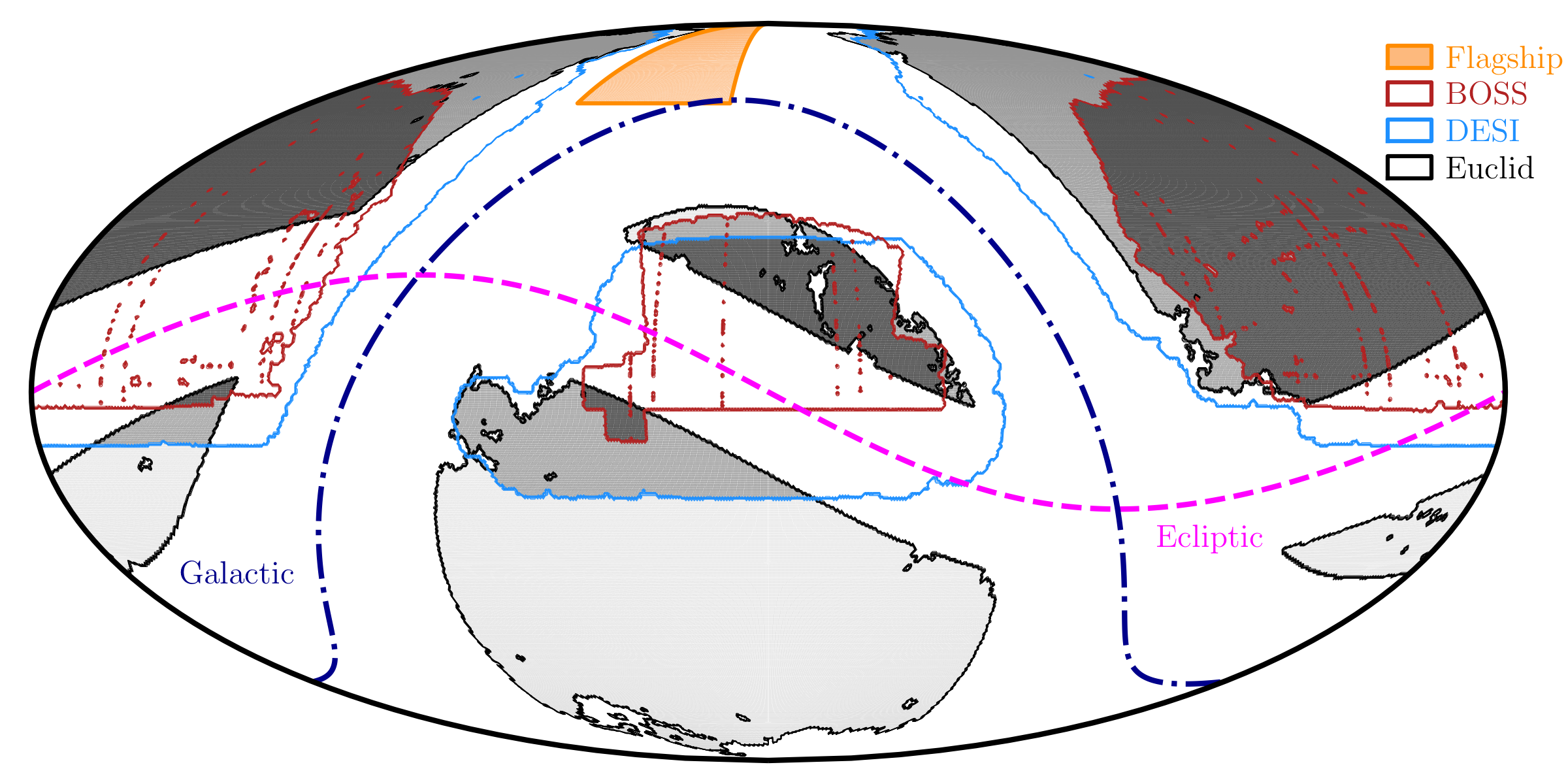}
	\caption{Survey footprints for BOSS, DESI, and \Euclid shown on the sky, with darker greys indicating regions of tracer--target sample overlap. The footprint perimeter for BOSS is indicated in red, for DESI in blue, and for \Euclid in black. Light grey indicates regions of the \Euclid survey alone, mid-grey indicates regions with \Euclid overlapping with either BOSS or DESI, and dark grey indicates regions with \Euclid overlapping both BOSS and DESI. For comparison, the on-sky subregion of the Flagship octant for which we have simulated photo-$z$ is shown in orange. The \Euclid survey region is defined to avoid both the Ecliptic (dashed pink line) and Galactic (dash-dotted blue line) planes.}
	\label{fig:euclid_footprints}
\end{figure*}

The footprints for the BOSS\footnote{\href{http://www.sdss3.org/dr9/algorithms/boss\_tiling.php\#footprint}{http://www.sdss3.org/dr9/algorithms/boss\_tiling.php\#footprint}}
, DESI,\footnote{\href{https://www.legacysurvey.org}{https://www.legacysurvey.org}}
and \Euclid surveys \citep{Amiaux2012, EuclidWide} are shown on the sky in Fig. \ref{fig:euclid_footprints}. Light grey areas indicate regions of the \Euclid survey with no overlap, darker grey indicates regions overlapping with at least one of BOSS or DESI, and the darkest grey indicates regions overlapping with both BOSS and DESI. For comparison we also indicate the footprint of the Flagship galaxy simulation used in this study, which is a subset of the full Flagship octant. \Euclid is projected to overlap with DESI over $9015\,{\rm deg}^{2}$, and with both BOSS and DESI over $6005\,{\rm deg}^{2}$. Unfortunately most of this overlap takes place across the Northern Hemisphere, where photometry from LSST will be limited to the southern-most regions, and therefore the accuracy of the photo-$z$ is likely be worse than that for the ones used in this study. Since these photo-$z$ catalogues were not produced without LSST-like photometry, we could not isolate this study to purely \Euclid photometry. We have instead chosen to assume that either the $ugrizy$ bands of LSST will be supplemented in the Northern Hemisphere by other photometric surveys of equivalent depth \citep[as assumed by ][]{Pocino2021} or that the spectroscopic overlaps of BOSS and DESI are supplied by an equivalent Southern Hemisphere survey such as those that will be observed with the 4-metre Multi-Object Spectroscopic Telescope\footnote{\href{https://www.4most.eu/cms/}{https://www.4most.eu/cms/}} (4MOST). This means the quality of the \Euclid photo-$z$ will depends on the angular footprint of the supplementary photometry used from other surveys such as LSST and needs to be factored into future studies as this may also result in $n(z)$ that are angular dependent -- which, if this is the case, will drastically complicate cosmological inference and analysis.

\subsection{\Euclid photometric sample}
\label{sec_photoz}

The Flagship simulation \citep{Potter2017} is a large $N$-body simulation computed with over two trillion dark matter particles in a box of length $3.78\, h^{-1}{\rm Gpc}$ using the \texttt{PKDGRAV3} $N$-body code. The simulation is constructed assuming a flat $\Lambda$ cold dark matter model with cosmological parameters $\Omega_{\rm m}=0.319$, $\Omega_{\rm b}=0.049$, $\Omega_{\rm \Lambda}=0.681$, $\sigma_{8}=0.83$, $n_{\rm s}=0.96$ and $h=0.67$. Dark matter haloes were identified using \texttt{ROCKSTAR} \citep{rockstar}, and galaxies assigned using halo abundance matching and halo occupation distribution techniques. Simulated \Euclid photo-$z$ were constructed for galaxies in the Flagship simulation over the RA range $15^{\circ}$ to $75^{\circ}$, and Dec range $62^{\circ}$ to $90^{\circ}$ \citep{Pocino2021}. We briefly describe the construction of this catalogue below but refer the reader to \citet{Pocino2021} for a complete description.

To generate realistic photo-$z$ catalogues, realistic fluxes were created by adding Gaussian noise to Flagship galaxy fluxes. The Gaussian noise is defined with a standard deviation of $f_{5\sigma}/5$ (the flux error associated with each galaxy) where $f_{5\sigma}$ is the limiting flux depth at a signal-to-noise ratio (S/N) of 5. The limiting magnitudes at $5\sigma$ depth considered to compute flux observations were 26.3, 27.5, 27.7, 27, 26.2, and 24.9 for the $ugrizy$ bands of Rubin LSST, and 24.6, 23, 23, and 23 for the $I_{\scriptscriptstyle\rm E}$ and $Y_{\scriptscriptstyle\rm E}$, $J_{\scriptscriptstyle\rm E}$, $H_{\scriptscriptstyle\rm E}$ bands of \Euclid at $10\sigma$ depth assuming both surveys are at end-of-survey depth.

The noisy $I_{\scriptscriptstyle\rm E}$ magnitudes were cut to less than $24.5$ \citep{Laureijs2011} to simulate the \Euclid $I_{\scriptscriptstyle\rm E}$ galaxy sample. Realistic estimates of the photometric redshift of simulated galaxies were made using the directional neighbourhood fitting (DNF; \citealt{Vicente2016}) training-based algorithm. The DNF photo-$z$ are estimated according to the proximity in colour-magnitude space of target and training galaxies, where redshifts are known for the training set. The DNF algorithm produces two photo-$z$ estimates; `$z_{\rm mean}$' takes the mean of neighbouring galaxies in colour and magnitude space, and `$z_{\rm mc}$' draws a random redshift from the neighbouring galaxies. A $3.35\,{\rm deg}^{2}$ patch of Flagship was used to create training samples for the DNF. The first training sample was defined to be fully representative in redshift and magnitude space, and produced the photo-$z$ catalogues $z_{\rm mean}$ and $z_{\rm mc}$. A second training sample featured a completeness drop in $I_{\scriptscriptstyle\rm E}$ magnitude (emulating the expected spectroscopic completeness fraction versus photometric depth function for surveys such as Rubin; see \citealt{Newman2015}), and produced the photo-$z$ catalogues $z_{\rm mean\, Rubin}$ and $z_{\rm mc\, Rubin}$. We repeated our analysis on each of these four photo-$z$ catalogues.

We produced ten tomographic redshift bins for each photo-$z$ catalogue. These were constructed by selecting galaxies with photo-$z$ between $0.2$ and $1.6$, and then dividing them into ten bins with approximately equal numbers of galaxies. The definition for the tomographic bins used in this study differ in two important ways to the definitions currently planned for \Euclid; in the real survey the maximum photo-$z$ will extend up to $z=2.6$ rather than $z=1.6$ and secondly 13 rather than 10 tomographic bins will be used \citep{Euclidphotoz2020}. The justification for these difference is motivated by two factors, (1) a limit to the galaxies in the Flagship mocks of $z<2.2$ and (2) a sharp cut off in simulated spectroscopic tracers to $z<1.8$. Including galaxies with photo-$z$ of greater than $z>1.6$ will introduce a substantial set of galaxies with true redshifts beyond $z>1.8$ where clustering redshift measurement cannot be made. Since this limitation is purely based on our setup we wish to limit any effect it may pose on the results and implications for \Euclid. Quasars (e.g. as observed by DESI; \citealt{Aghamousa2016}) will eventually provide reference samples for calibration at these high redshifts. We note that the clustering redshifts method will need to be carefully optimised for quasar samples, where bias modelling and angular scale utilisation are likely to pose challenges for the much sparser quasar population.

We note that \textit{Euclid}'s ground-based photometry complement in the Northern Hemisphere will be somewhat shallower and less homogeneous than the LSST. Nevertheless, the photometry will meet the stringent Euclid requirements and hence yield high-quality photometric redshifts, so that our conclusions remain valid.

\subsection{Reference spectroscopic samples}
\label{specsample}

We detail below the various spectroscopic tracer samples that we define in the Flagship simulation, each designed to mimic a current or future galaxy sample observed by \Euclid, BOSS, or DESI.

\subsubsection{\Euclid NISP-S samples}

The \Euclid near-infrared spectrometer \citep{EuclidNISP} is designed to measure spectroscopic redshifts for over 50 million galaxies (referred to as the NISP-S sample). The redshift determination is to be made with slitless spectroscopy dependent on the detection of emission lines, in particular the H$\alpha$ line, in the near-infrared. We defined mock NISP-S samples in Flagship by selecting galaxies with H$\alpha$ fluxes greater than $3\times10^{-16}$ ${\rm erg}\, {\rm cm}^{-2}\, {\rm s}^{-1}$ \citep{Laureijs2011}, and cutting to a redshift range between $0.9$ and $1.8$. Comparisons of the mock NISP-S redshift distribution to the optimistic and pessimistic predictions made by \citet{EuclidPrep2020}, we found consistency above redshifts of $1.35$, but an overproduction of galaxies at lower redshifts. To correct for this we carry out sparse sampling for mock NISP-S galaxies with $z<1.35$ to approximately match the optimistic expectation from \citet{EuclidPrep2020}.

\subsubsection{Baryon Oscillation Spectroscopic Survey LOWZ \& CMASS samples}

We defined BOSS-like samples according to the colour-magnitude selections specified by \cite{Dawson2013}. BOSS produced two spectroscopic LRG samples, LOWZ and CMASS, targeting adjacent redshift intervals $z\in[0.15,0.43]$ and
$z\in(0.43,0.7],$ respectively, and with true densities of about $30\,{\rm deg}^{-2}$ and $120\,{\rm deg}^{-2}$, respectively. We replicated the BOSS colour-magnitude selections as follows, first defining
\begin{align}
& c_{\parallel} = 0.7\,(g-r) + 1.2\,(r-i-0.18)\:,\\
& c_{\perp} = (r-i) - (g-r)/4 - 0.18\:,\\
& d_{\perp} = (r-i) - (g-r)/8\:,
\end{align}
where $gri$ denote apparent magnitudes (at our simulated LSST depth). We then defined mock LOWZ samples with the following criteria:
\begin{align}
& |c_{\perp}| < 0.2\:,\\
& r < 13.6 + c_{\parallel}/0.3\:,\\
& 16 < r < 19.5\:,
\end{align}
and mock CMASS samples with
\begin{align}
& 17.5 < i < 19.9\:,\\
& d_{\perp} > 0.55\:,\\
& i < 19.86 + 1.6\,(d_{\perp} - 0.8)\:.
\end{align}
These cuts provided LOWZ- and CMASS-like objects in the Flagship simulation at rates of $67\,{\rm deg}^{-2}$ and $139\,{\rm deg}^{-2}$, respectively. We sparse sample our selections to the desired densities, with sparse sampling fractions of $0.44$ and $0.86$.

\subsubsection{Dark Energy Spectroscopic Instrument BGS, LRG, and ELG samples}

We based our DESI-like sample selections on those detailed in Sect. 3 of \cite{Aghamousa2016}. DESI will measure the spectra of galaxies at relatively low redshifts with the BGS (designed to use spectra obtained during brighter sky conditions), and will also target two deeper surveys of LRGs and ELGs.

We define our mock BGS sample with the following criteria,
\begin{align}
& r<19.5\:,\\
& z < 0.4\:,
\end{align}
where $z$ is the true object redshift. We define the mock LRG sample with
\begin{align}
& z < 20.46\:,\\
& r < 23\:,\\
& 0.6 < z < 1.0\:,\\
& \texttt{color\_kind}  = \rm{red\,sequence} \, (0)\:,
\end{align}
where \texttt{color\_kind} is a flag labelling red sequence (0), green valley (1) and blue cloud (2) galaxies. Lastly, we defined the mock ELG sample of galaxies with {\sc [O ii]} emission line strengths greater than $8\times10^{-17}\,\rm{erg\,s^{-1}\,cm^{-2}}$, and with the following cuts,
\begin{align}
& r < 23.4\:,\\
& g - r < 0.7\:,\\
& r - z > 0.3\:,\\
& 0.6 < z < 1.6\:,\\
& \texttt{color\_kind} \neq \rm{red\,sequence} \, (0)\:,
\end{align}
which act to isolate ELGs in colour-colour space, minimising contamination of the sample by lower-redshift objects or by stars. We also applied a hard cut to the expected redshift range, and required that the \texttt{color\_kind} not specify a red sequence object, though these have comparatively small impacts upon the selection.

These cuts provided BGS-, LRG-, and ELG-like objects in the Flagship simulation at rates of $1174\,{\rm deg}^{-2}$, $392\,{\rm deg}^{-2}$, and $2021\,{\rm deg}^{-2}$, respectively. We achieved the expected number densities (of $700\,{\rm deg}^{-2}$ for BGS, $285\,{\rm deg}^{-2}$ for LRG, and $1220\,{\rm deg}^{-2}$ for ELG) by sparse sampling with sparse sampling fractions of $0.6$, $0.73$, and $0.6,$ respectively.

We omitted DESI LRGs with $z<0.6$, which have selection criteria similar to BOSS. Therefore, the dearth of reference galaxies in the range $0.4<z<0.6$ will be much less pronounced for real data, which makes our analysis conservative.

\section{Method}
\label{sec:method}

In this section we briefly outline the clustering redshift method used in this study, the subsequent method for modelling the redshift distributions, and lastly the area rescaling approach used to predict the uncertainty on the mean redshift for the full \Euclid survey.

\subsection{Clustering redshifts}
\label{sec:cc_theory}
The angular correlation function $\omega_{\rm 12}(\theta, z)$ for two samples on the sky (denoted by 1 and 2), where sample 1 is at a fixed redshift $z$, is given by the \citet{Limber1953} relation,
\begin{equation}
	\omega_{12}(\theta, z)=b_{1}(z)\,\int_{0}^{\infty}{\rm d}z'\,n_{2}(z')\,b_{2}(z')\,\xi\,\Big[R(\theta,z, z'), z\Big]\:,
	\label{eq:limber_crosscorrelation}
\end{equation}
where $b_{1}(z)$ and $b_{2}(z)$ are the redshift-dependent biases for samples 1 and 2, respectively, $n_{2}(z)$ is the normalised redshift selection function for sample 2, $\xi(R, z)$ is the matter auto-correlation function at redshift $z$ and comoving three-dimensional separation
\begin{equation}
	R(\theta, z, z') = \sqrt{\Big[\chi(z)-\chi(z')\Big]^{2}+\Big[f_{\rm K}(z')\,\theta\Big]^{2}}\:,
\end{equation}
where $\chi(z)$ is the radial comoving distance at redshift $z$ and $f_{\rm K}(z)$ is the angular diameter distance at redshift $z$.

For the computation of clustering redshifts, the following assumptions are made. We first consider the angular correlation function for a spectroscopic tracer sample (denoted with an index $\rm s$) and photometric target sample (denoted with an index $\rm p$)
\begin{equation}
	\omega_{\rm sp}(\theta, z)=b_{\rm s}(z)\,\int_{0}^{\infty}{\rm d}z'\,n_{\rm p}(z')\,b_{\rm p}(z')\,\xi\,\Big[R(\theta,z, z'), z\Big]\:.
	\label{eq:limber_crosscorrelation2}
\end{equation}
Angular correlation functions are computed for narrow redshift slices of the spectroscopic tracer with a mean redshift $z_{i}$ (where $i$ denotes the redshift bin) and bin width of $\Delta z$. Furthermore, assuming measurements are only made for small values of $\theta$ we can can make the approximation that $\xi\neq0$ only if the integral is calculated within the bin's redshift range $z_{i}\pm \Delta z/2$. The narrow sizes of the bins means $n_{p}$ and $b_{p}$ can be approximated as constants and therefore Eq. (\ref{eq:limber_crosscorrelation2}) simplifies to
\begin{equation}
	\omega_{\rm sp}(\theta, z_{i})=b_{\rm s}(z_{i})\,n_{\rm p}(z_{i})\,b_{\rm p}(z_{i})\,\omega_{ii}(\theta, z_{i})\:,
	\label{eq:limber_crosscorrelation3}
\end{equation}
where
\begin{equation}
	\omega_{ii}(\theta, z_{i})=\int_{z_{i}-\Delta z/2}^{z_{i}+\Delta z/2}{\rm d}z'\,\xi\,\Big[R(\theta,z_{i}, z'), z_{i}\Big]\:.
\end{equation}
In our study these expressions were evaluated for the transverse physical scale $r=\theta\chi/(1+z)$, which led to the final definition of the angular correlation function for clustering redshifts,
\begin{equation}
	\omega_{\rm sp}(r, z_{i})=b_{\rm s}(z_{i})\,n_{\rm p}(z_{i})\,b_{\rm p}(z_{i})\,\omega_{ii}(r, z_{i})\:.
	\label{eq:limber_crosscorrelation4}
\end{equation}
Thus we can evaluate the target sample's redshift selection function $n_{\rm p}$ by rearranging Eq. (\ref{eq:limber_crosscorrelation4}) for
\begin{equation}
	n_{\rm p}(z_{i}) = \frac{\omega_{\rm sp}(r, z_{i})}{b_{\rm s}(z_{i})\,b_{\rm p}(z_{i})\,\omega_{ii}(r, z_{i})}\:.
	\label{eq:cc1}
\end{equation}
The above equation requires knowledge of the bias-redshift relation for both the spectroscopic tracer and photometric target sample. By evaluating Eq. (\ref{eq:limber_crosscorrelation4}) for the angular auto-correlation functions for the spectroscopic and photometric samples we can define the bias functions as
\begin{equation}
	b_{\rm x}= \sqrt{\frac{\Delta z\,\omega_{\rm xx}(r, z_{i})}{\omega_{ii}(r, z_{i})}}\:,
\end{equation}
where $\rm x$ can denote either the spectroscopic tracer or photometric target sample. Since the auto-correlation function is evaluated in a single, narrow redshift bin, the normalised redshift selection function $n_{\rm x}\rightarrow1/\Delta z$. Using these relations, Eq. (\ref{eq:cc1}) can be expressed as
\begin{equation}
	n_{\rm p}(z_{i}) = \frac{\omega_{\rm sp}(r, z_{i})}{\Delta z \,\sqrt{\omega_{\rm ss}(r, z_{i})\,\omega_{\rm pp}(r, z_{i})}}\:.
	\label{eq:cc2}
\end{equation}
However, in practice $\omega_{\rm{pp}}(r,z)$ is difficult to obtain, since we do not know the photometric sample redshifts {a priori}, and thus cannot bin target galaxies to measure the correct auto-correlations and anchor the photometric sample galaxy bias. The result is that the $n_{\rm{p}}(z)$ is biased and requires a correction scheme to account for any redshift evolution of the photometric sample galaxy bias. Several bias mitigation strategies can be employed.

Method 1: No corrections are applied. This is useful to test the success of the following bias correction methods. This means we assume $n_{\rm p}(z) \propto \omega_{\rm sp}(r,z)$.

Method 2: The spectroscopic biases are computed from the auto-correlation functions and are incorporated in the $n_{\rm p}(z)$ computation. This means we use Eq.~(\ref{eq:cc2}) but $\omega_{\rm pp}(r,z)$ is evaluated for the entire tomographic bin instead of within thin redshift slices.

Method 3: `Self-consistent bias mitigation' uses a one-parameter redshift power law, $\mathcal{B}_{\alpha}(z),$ fitted to the observed auto- and cross-correlations to account for redshift evolution in $b_{\rm p}$ \citep{Davis2018, Busch2020}
\begin{equation}
	\mathcal{B}_{\alpha}(z) = (1 + z)^{\alpha} \propto \sqrt{\Delta z_{i}\,\bar{\omega}_{\rm{pp}}(z)}\:,
	\label{eq:bias_powerlaw_form}
\end{equation}
where $\alpha$ is a free model parameter, and barred correlations $\bar{\omega}(z)$ correspond to integrals over $\omega(r,z)$ between chosen limits $r_{\rm{min}}$ and $r_{\rm{max}}$. This means $\omega_{\rm pp}$ in Eq.~(\ref{eq:cc2}) is replaced with $\mathcal{B}_{\alpha}(z) ^{2} /\Delta z$.

Method 4: The bias for the photometric sample is computed from the auto-correlation function for only the photometric galaxies with true redshifts within a given redshift slice. Eq. (\ref{eq:cc2}) is therefore applied exactly. We note that this method can only be applied to simulated data, and is useful for assessing the accuracy of other bias correction schemes.

\subsection{Cross-correlating redshifts}
\label{sec:cc_implement}

We compute clustering correlations in the Flagship data using the single-bin (defined by $r_{\rm{min}},r_{\rm{max}}$) method of \citet{Schmidt2013} and \citet{Menard2013}, implemented with the Davis-Peebles clustering estimator \citep{Davis1983} as follows:
\begin{equation}
	\bar{\omega} = \frac{N_{\rm R}\int^{r_{\rm{max}}}_{r_{\rm{min}}}{\rm{d}}r \, W(r) \, {\rm DD}(r)}{N_{\rm D}\int^{r_{\rm{max}}}_{r_{\rm{min}}}{\rm{d}}r \, W(r) \, {\rm DR}(r)} - 1\:,
\end{equation}
where $W(r)\propto{}r^{\beta}$ (e.g. $\beta=-1$; \citealt{Schmidt2013,Menard2013}), ${\rm DD}(r)$ and ${\rm DR}(r)$ are galaxy-galaxy and galaxy-random pair counts (in the comoving bin centred on $r$), respectively, and $N_{\rm R}/N_{\rm D}$ re-normalises pair counts to account for over-sampling of random points relative to the data. The method follows \citet{Busch2020} and uses the software package \texttt{YAW}\footnote{\url{https://github.com/jlvdb/yet_another_wizz}}  described in \citet{Busch2020} with errors and covariance matrices estimated using bootstrap resampling.

\subsection{Redshift distribution fitting}

Since the clustering redshift measurements have Gaussian error distributions, they allow for negative $n(z)$ values. Of course, no true PDF can exhibit this property and as a result the raw clustering redshift measurements will artificially inflate the uncertainties on the mean redshift $\langle z\rangle$. Therefore, we did not determine $\langle z\rangle$ directly from the clustering redshift measurements but rather by fitting model $n(z)$ distributions. Two model $n(z)$ distributions were considered. The first model, the shifted-true model, is simply the true $n(z)$ measured from simulations with the addition of two free parameters: $A$ and $\delta z$, which are the amplitude and shift parameters, respectively (explained in greater detail in Sect. \ref{shiftedtruemodel}). The second model is the suppressed Gaussian process (suppressed-GP) model discussed in Sect. \ref{gpmodel}, which, unlike the first model, makes no assumptions about the shape of the $n(z)$ distribution. In both cases, the models are not normalised, since normalisation cannot be enforced on the clustering redshift measurements. Normalisation of the model is therefore only conducted when we are calculating the mean of the PDF.

\subsubsection{Shifted-true}
\label{shiftedtruemodel}

In the shifted-true model the clustering redshift measurements are fitted to the true $n(z)$. If these were real data, this true $n(z)$ would not be known and would have to be estimated through forward modelling on simulations such as Flagship, or from direct calibration \citep{Busch2020}. The measured $n(z)$ is fitted with two parameters, a shift along the redshift direction $\delta z$ and an amplitude $A$,
\begin{equation}
n_{\rm model}(z; \delta z, A) = A\,n_{\rm true}(z+\delta z)\:.
\end{equation}
Fitting with the true distribution allows us to test the robustness of the output clustering redshift measurements, since the resultant best-fit values should be consistent with $\delta z=0$ in the absence of systematic errors. The parameters are determined using a Gaussian likelihood,
\begin{equation}
\ln\mathcal{L}\,\big[\delta z, A\, |\, \hat{n}(z), \sigma_{\hat{n}}(z)\big] = -\frac{1}{2}\left[\frac{\hat{n}(z)-n_{\rm model}(z; \delta z, A)}{\sigma_{\hat{n}}(z)}\right]^{2}\:,
\end{equation}
where $\hat{n}(z)$ and $\sigma_{\hat{n}}(z)$ are the clustering redshift measurements and 68.3\% confidence intervals, respectively.

\begin{figure}
	\centering
	\includegraphics[width=\columnwidth]{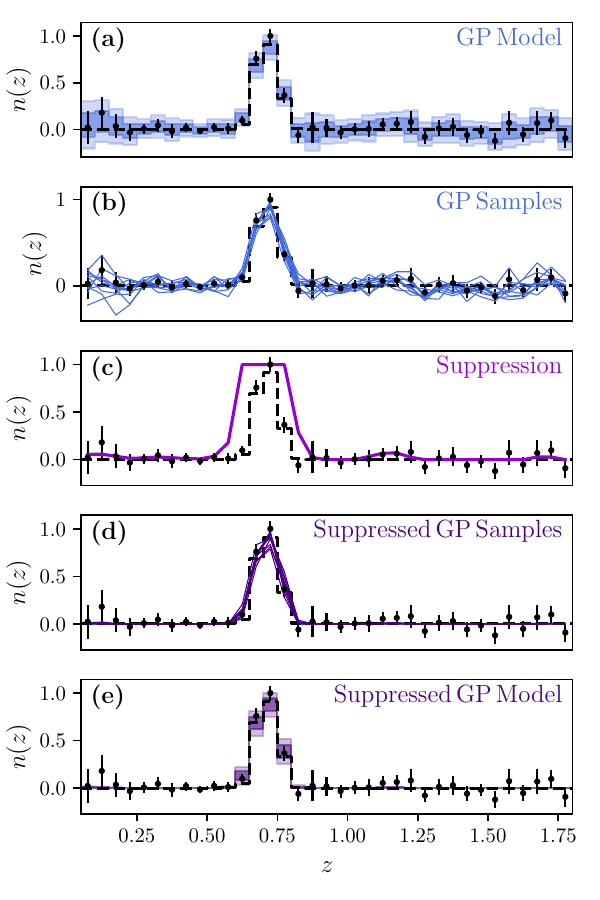}
	\caption{Method for fitting suppressed-GP models to clustering redshift, $n(z),$ measurements. The true $n(z)$ for one tomographic bin is shown with dashed black lines. The measured clustering redshift, $n(z),$ is shown with black markers and error bars. \emph{(a)} We fit a GP model to the clustering redshift, $n(z),$ measurements. \emph{(b)} We draw random realisations from the GP. \emph{(c)} We construct a suppression function, taking as input the GP draw and the smoothed S/N of the clustering redshift measurement. \emph{(d)} We multiply the random realisations from the GP by the suppression function. This ensures that the suppressed-GP model is always positive and suppresses low-S/N fluctuations in the GP. \emph{(e)} We construct 68.3\% and 95.5\% confidence envelopes from samples of the suppressed-GP model.}
	\label{fig:suppress_gp}
\end{figure}

\begin{figure*}
	\centering
	\includegraphics[width=17cm]{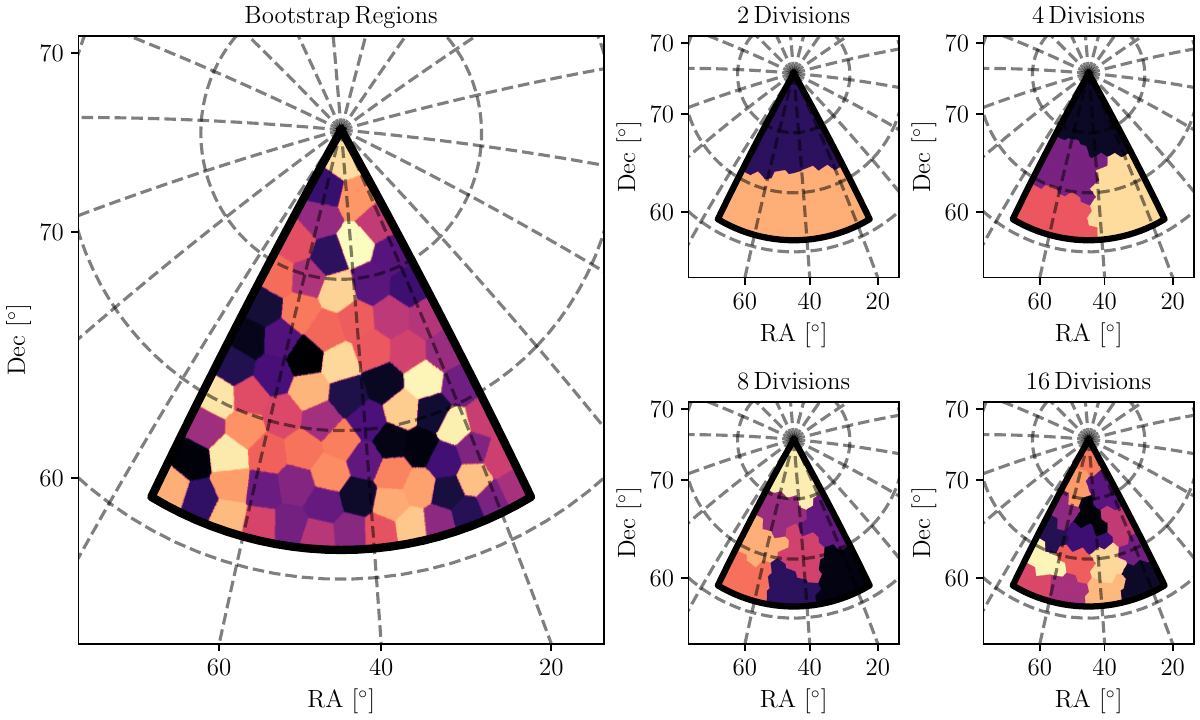}
	\caption{Flagship footprint and area rescaling subregions. \emph{Left}: Bootstrap subregions, indicated with different colours, that are used to calculate errors in \texttt{YAW}. \emph{Right}: Different subregions of the Flagship footprint, shown with different colours, used to establish the relation between the tracer--target overlap area and the uncertainty on the mean redshifts of tomographic bins. The four subpanels show the Flagship footprint divided into 2 (\emph{top left}), 4 (\emph{top right}), 8 (\emph{bottom left}), and 16 (\emph{bottom right}) subregions. These are plotted on an orthographic projection of a subset of the sky focused on the Flagship footprint. The RA and Dec grid is shown with dashed grey lines. The values of the grid lines are indicated where they intercept the x-axis for the RA grid and the y-axis for the Dec grid.}
	\label{fig:bootstraps_area}
\end{figure*}

\begin{figure*}
	\centering
	\includegraphics[width=17cm]{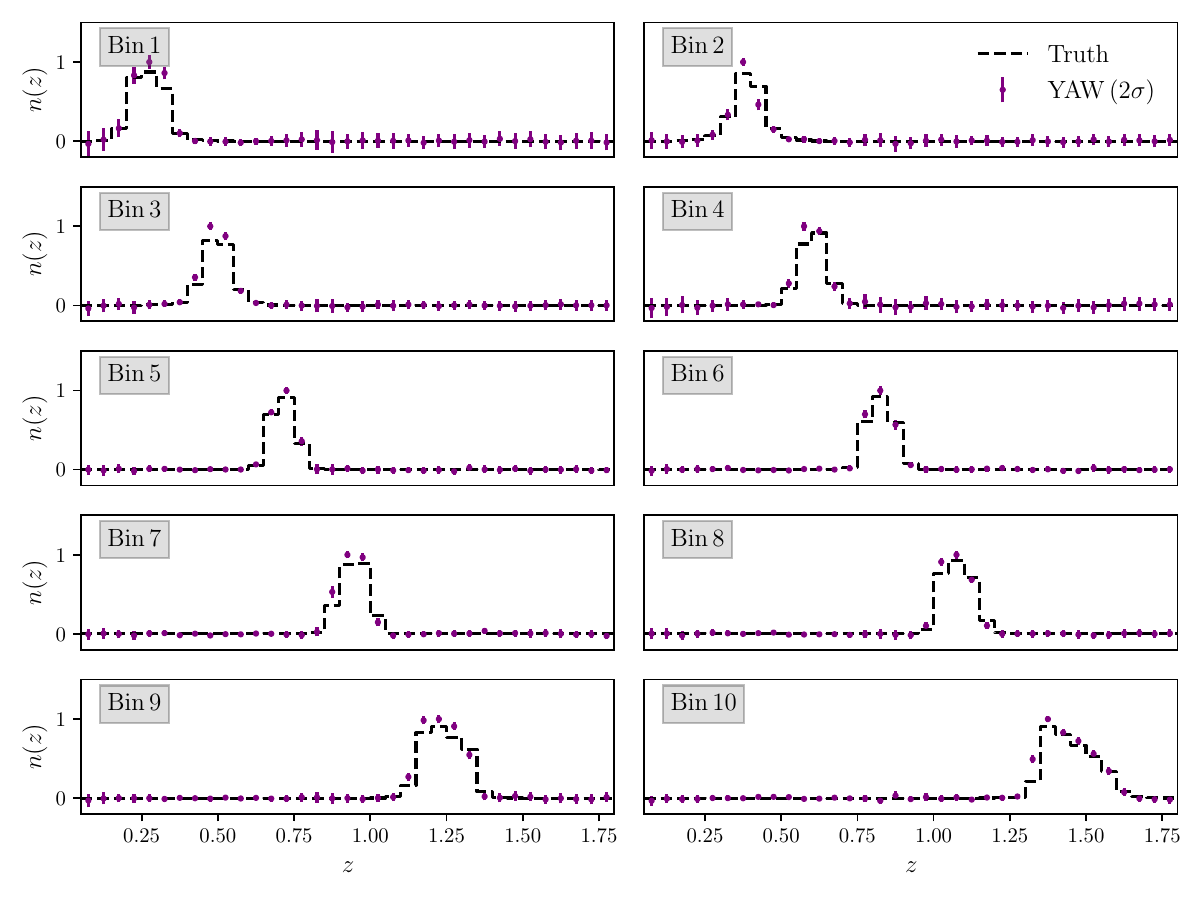}
	\caption{Clustering redshift measurements in comparison to the true $n(z)$ for ten simulated \Euclid Flagship tomographic photo-$z$ bins. Measurements from \texttt{YAW} are shown in purple, with error bars indicating 95.5\% confidence intervals. The measurements clearly trace the true $n(z)$, albeit with some spurious local fluctuations. These fluctuations probably arise due to an incomplete galaxy bias correction methodology. The bias correction method used in this figure is method 2 (Sect. \ref{sec:cc_theory}).}
	\label{fig:cc_nz}
\end{figure*}

\subsubsection{Suppressed Gaussian process}
\label{gpmodel}

Whilst the shifted-true model is a valuable test for systematic errors, it is important to note that we will not have access to such a model for real data. In previous studies, clustering redshifts were fitted to models based on simulated photometric redshift samples while \citet{GattiDES2021} used direct calibration of the data. Although the simulated photometric redshift $n(z)$ may work as a good proxy for the real model, it may be better to have a model that is more flexible and could be relied upon in a more general setting. With this in mind, we present a new, non-parametric approach to fitting clustering redshift distributions based upon Gaussian processes (GPs, and featuring a suppression function that damps signals in regions where the clustering redshift measurements are consistent with zero.

A GP is first fitted to the clustering redshift measurements. This is carried out by the \texttt{Python} package \texttt{George}\footnote{\url{https://george.readthedocs.io/en/latest/}} using a Matern-3/2 kernel. One benefit of such a model is that we can now draw random samples from the distribution to measure the mean redshift and its uncertainty. However, as with the clustering redshift measurements there is nothing limiting the GP model from drawing samples that are negative. This issue is further compounded by the fact that uncertainties crossing zero will create spurious fluctuations in the GP samples. To ensure that the GP is positive, and to remove spurious signals where the clustering redshift measurement is consistent with zero, we apply a suppression function to the GP realisations. The suppression function is defined by the following expression,
\begin{equation}
	S(x,k) =\begin{dcases}
		0\,, &x\leq 0\:,\\
		1 - (1-x)^{k}\,, & 0<x<1\:,\\
		1\,, &x\geq 1\:,
	\end{dcases}
	\label{suppression_function}
\end{equation}
where $k$ is a damping factor taken to be $0.3$ in this study and
\begin{equation}
	x = \frac{ n_{\rm GP}^{\rm i}(z) }{ \Sigma_{\rm T}\,\Big( \mathcal{G}*\Sigma \Big) }\:,
\end{equation}
$n_{\rm GP}^{\rm i}(z)$ is a random GP realisation, $\Sigma_{\rm T}=3$ is the S/N threshold, $\Sigma=n_{GP}(z)/\sigma_{GP}(z)$ is the S/N function (where $n_{\rm GP}$ and $\sigma_{\rm GP}$ are the GP mean and 68.3\% confidence interval), $\mathcal{G}$ is a Gaussian with standard deviation of $0.05$ and $*$ is the convolution operator. The Gaussian convolution with the S/N prevents spurious random fluctuations in the S/N from impacting the activation of the suppression function and ensures the tails of the clustering redshift $n(z)$ are not too harshly damped. A standard deviation drastically smaller than $0.05$ will be smaller than the bin size of the clustering redshift measurements and therefore will be equivalent to no smoothing. On the other hand, larger values will more strongly correlate bins that may be problematic since, in some cases, the $n(z)$ profile spans across only $\sim$5 bins and may be completely washed out.

An example of this procedure is shown in Fig. \ref{fig:suppress_gp}. The suppressed-GP has the desired effect, that is to say, noisy PDFs are not drawn in regions of the clustering redshifts that are consistent with zero. A potential shortcoming is that tails in the distribution will be suppressed if they have low S/N. Steps to mitigate this effect could be studied in future work by exploring alternative suppression functions and techniques.

\subsection{Area rescaling}

We seek to provide estimates for the uncertainties of the $n_{\rm{p}}(z)$ determination as a function of the projected \Euclid data volume; thus we must extrapolate from the $402\,{\rm deg}^{2}$ of the Flagship area to the larger spectroscopic overlap areas, and also attempt to characterise the impact of sample variance. \Euclid is projected to overlap with DESI over $9015\,{\rm deg}^{2}$, and with both DESI and BOSS over $6005\,{\rm deg}^{2}$. To rescale the determined uncertainties on mean redshifts, we divided the Flagship lightcone into multiple subdivisions (see Fig. \ref{fig:bootstraps_area}) and re-ran our analysis on each of these subdivisions independently. This allowed us to approximate the relationship between the tracer--target overlap area and the uncertainties on $\langle z \rangle$. We then fitted these relationships with power laws, and extrapolated to estimate the uncertainties from the full \Euclid, BOSS, and DESI overlap area. We note we do not redefine the bootstrap regions when we run the analysis on each subdivision, this means estimates of the error are less accurate for smaller regions since they are based on fewer defined bootstrap regions. An alternative approach would be to use purely shot-noise covariance, but we leave this for future work. Furthermore, since these regions overlap, measurements on the subregions will be strongly correlated.

\section{Results}
\label{sec:results}

This section details the results of our clustering redshift calibration, focusing on the uncertainties of mean redshift determination across our ten mock tomographic bins.

\subsection{Clustering redshift measurements}

We measured clustering redshifts using the \texttt{YAW} software package between scales $r_{\min}=100\,{\rm kpc}$ and $r_{\max}=1000\,{\rm kpc}$ \citep[following the scales used by KiDS;][]{Busch2020}, considering separately each of the four simulated \Euclid photometric redshift catalogues presented by \citet{Pocino2021}. We performed the analysis independently for the full Flagship region (with photo-$z$), and for each subregion (shown in Fig. \ref{fig:bootstraps_area}).

In Fig. \ref{fig:cc_nz} we compare the clustering redshift measurements obtained by \texttt{YAW} (shown with purple error bars) to the true redshift distribution of the `mean Rubin' photometric catalogue on Flagship. The measurements clearly trace the true distribution, and showcase the capability of clustering redshifts to retrieve redshift distributions for photometric catalogues. There are, however, some significant fluctuations, for example at the peak of bin 2. The measurements in Fig. \ref{fig:cc_nz} are corrected for the spectroscopic sample galaxy bias (method 2, discussed in Sect. \ref{sec:cc_theory}).

\subsection{Mean redshifts}

\begin{figure*}
	\centering
	\includegraphics[width=17cm]{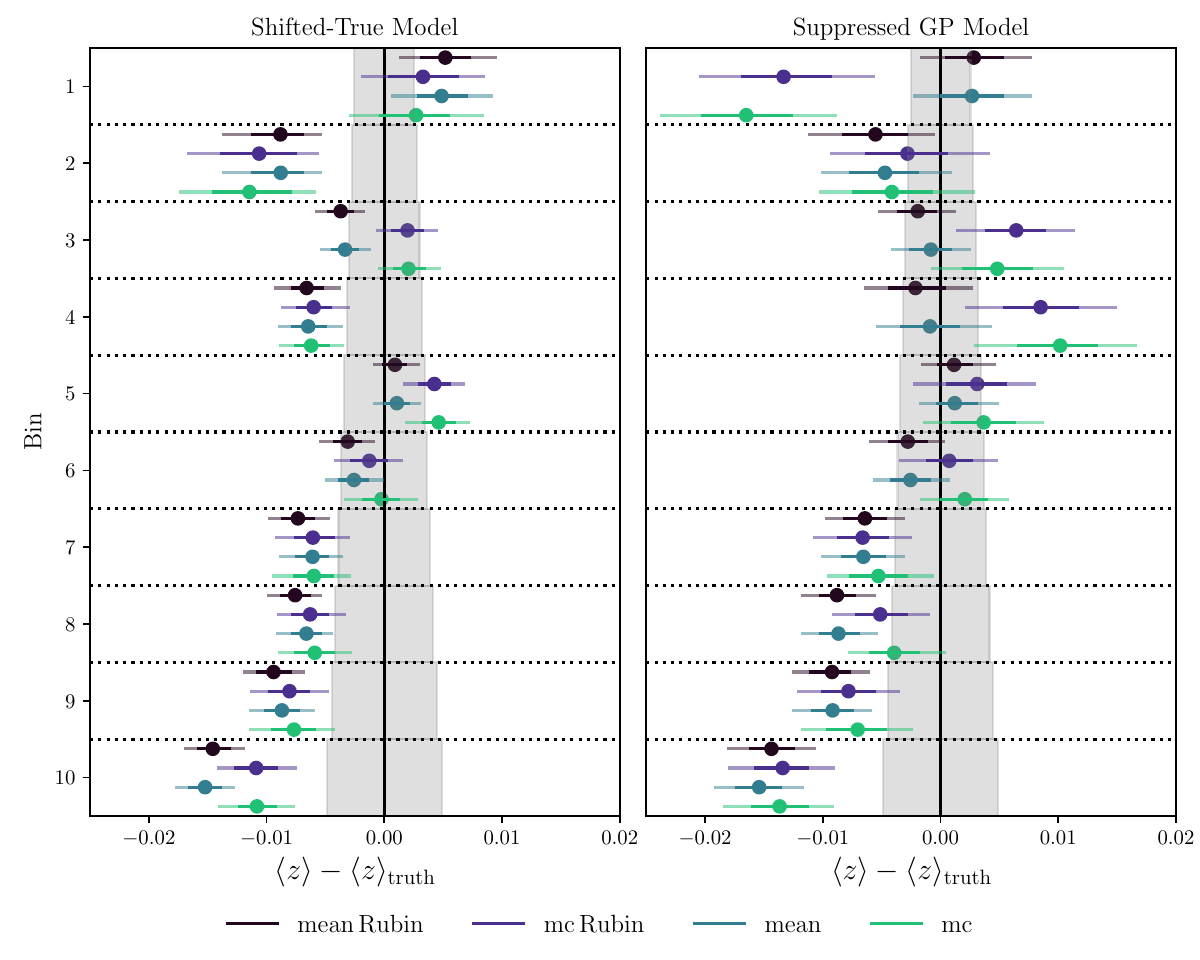}
	\caption{Error on the mean redshift determination $\langle z \rangle - \langle z \rangle_{\rm truth}$ for ten \Euclid Flagship tomographic bins, for each of our four photometric redshift catalogues (`mean Rubin' in dark purple, `mc Rubin' in purple, `mean' in dark green, and `mc' in light green, with dark and light shades indicating 68.3\% and 95.5\% uncertainties). The grey bands show the target uncertainties for \Euclid. Two $n(z)$ fitting methods are shown: the shifted-true model (\emph{left}, Sect. \ref{shiftedtruemodel}) and the suppressed-GP model (\emph{right}, Sect. \ref{gpmodel}), each fit to clustering redshift measurements made with spectroscopic sample bias corrections (i.e. method 2, Sect. \ref{sec:cc_theory}).}
	\label{fig:cc_redshift}
\end{figure*}

Having measured the clustering redshifts, we wished to constrain and compare each bin's $n(z)$ distribution by measuring the mean redshift $\langle z\rangle$ and comparing to the truth. The mean redshift is determined by calculating
\begin{equation}
	\langle z\rangle = \sum_{i=1}^{N_{\rm CC}}\,z_{i}\,n(z_{i})\,\Delta z\:,
	\label{eq:mean}
\end{equation}
where $n(z)$ is the normalised model redshift distribution within the redshift ranges $z_{\min}$ and $z_{\max}$, $z_{i}$ is the redshift centre of a clustering redshift bin and $n(z_{i})$ a clustering redshift measurement. The $i$ is used to denote a specific measurement and $N_{\rm CC}$ the total number of measurements. We computed $\langle z\rangle$ for the $n(z)$ model only within the redshift range where clustering redshift measurements were made (therefore $z_{\min}=0.05$ and $z_{\max}=1.8$) while true values $\langle z\rangle_{\rm truth}$ were computed from the true $n(z)$ measured across the entire Flagship range $z<2.2$. For both the shifted-true and suppressed-GP models, we drew 1000 realisations of the redshift distribution (for the shifted-true model, this means drawing random shift parameters from the posterior likelihood). We determine the mean redshift $\langle z\rangle$ from the sample $n(z)$, and the standard deviation on the mean $\sigma \left( \langle z\rangle \right)$ from the variance of the sample $\langle z\rangle$. In Fig. \ref{fig:cc_redshift} we display the mean redshift error $\langle z\rangle - \langle z\rangle_{\rm truth}$ for both models, in each bin and for the different photometric catalogues, constrained by the full Flagship footprint. The constraints show significant biases, in particular for bins 7--10, which are biased to lower redshifts for both models. For the shifted-true model we see some significant biases in bins 2 and 4, independent of photo-$z$ method. The suppressed-GP is biased for bins 1 and 4, but only for the photometric catalogue `mc'. The broadly similar performance of the suppressed-GP and shifted-true models shows that the suppressed-GP model is performing well, and that it is a robust alternative for $n(z)$ fitting. We also note that the directions and amplitudes of biases appear to be consistent between models, suggesting that systematic errors are independent of the fitting method for this analysis. Rather surprisingly, bin 3, which is placed in the region ($z=0.4$ to $0.6$) with fewer spectroscopic tracers, is relatively unbiased. One interesting property of this bin is that the tails on both sides are well constrained, due to the high number of spectroscopic tracers on either of the tails of the distribution. This suggests the clustering redshift performance on the tails may be driving biases in the other results. We note that we also find the errors on $z_{\rm mean}$ and $z_{\rm mean Rubin}$ to be smaller than the $z_{\rm mc}$ and $z_{\rm mc Rubin}$ for the suppressed-GP owing to the larger tails in the `mc' methods.

\begin{figure}
	\centering
	\includegraphics[width=\columnwidth]{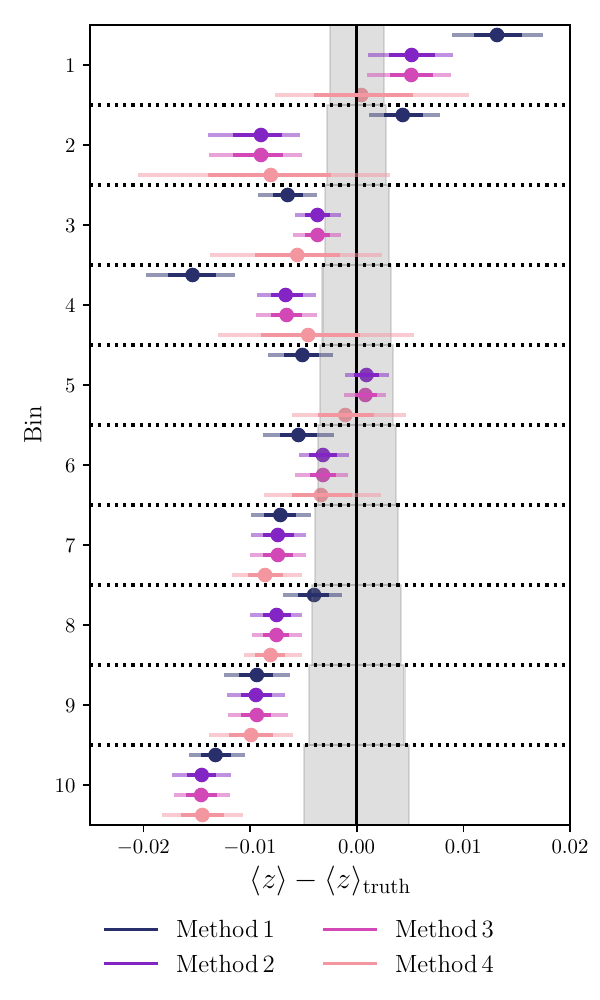}
	\caption{Error on the mean redshift determination, $\langle z \rangle - \langle z \rangle_{\rm truth}$, in ten \Euclid tomographic redshift bins, for each of the four galaxy bias correction strategies (Sect. \ref{sec:cc_theory}), compared via the shifted-true model (Sect. \ref{shiftedtruemodel}). In dark blue is method 1, where no corrections are applied; in purple is method 2, where the spectroscopic galaxy bias is calibrated by spectroscopic auto-correlations; in magenta is method 3, where the photometric galaxy bias is calibrated by a power-law fit to noisy, photo-$z$-binned photometric auto-correlations; and in light pink is method 4, where the photometric galaxy bias is calibrated using auto-correlations binned by true redshifts (not applicable to real data). Dark and light shades indicate 68.3\% and 95.5\% confidence intervals. The grey band shows the target uncertainties for \Euclid. Method 4 shows that systematic biases in bins 1--6 are caused by incomplete bias corrections, whilst the biases persist for higher-redshift bins, suggesting some other source of systematic error.}
	\label{fig:cc_redshift_method}
\end{figure}

In Fig. \ref{fig:cc_redshift_method} we compare the bias correction methods outlined in Sect. \ref{sec:cc_theory} by calculating the reduced $\chi_{\rm r}^{2}$ for each method,
\begin{equation}
	\chi_{\rm r}^{2} = \frac{1}{N_{\rm Bins}}\sum_{i=1}^{N_{\rm Bins}} \left[\frac{\langle z \rangle - \langle z \rangle_{\rm truth}}{0.002\,(1+\langle z \rangle_{\rm truth})}\right]^{2},
\end{equation}
on the photometric catalogue `mean Rubin', where $N_{\rm Bins}$ is the number of tomographic bins. We see that method 1 (no galaxy bias correction) performs worst with a $\chi^{2}_{\rm r}=7.8$, while method 2 (spectroscopic sample bias correction; the baseline method for all other figures) performs slightly better with a $\chi^{2}_{\rm r}=4.2$. Method 3, which fits a power law for photometric sample bias correction, performs very similarly to method 2 -- with a $\chi^{2}_{\rm r}=4.2$. Method 4, which performs the best with a $\chi^{2}_{\rm r}=3.7$, measures the true bias for photometric galaxies in each of the spectroscopic slices, and cannot be measured on real data. However, in simulations like Flagship, it allows us to test the significance and implications of residual biases. Method 4 shows that correcting for photometric galaxy biases in bins 1-6 removes any systematic biases on the determination of the mean redshift, and indicates that the mean redshift uncertainties may be underestimated by other methods. It should be possible to calibrate for such underestimations of $\sigma(\langle z \rangle)$ with realistic simulations like Flagship. We assess whether or not the extra noise component revealed by method 4 is likely to affect our conclusions, as we forecast uncertainties on $\langle z \rangle$ versus tracer--target overlap area in Sect. \ref{sec:area_rescaling_results}. Bins 7--10 remain significantly biased in $\langle z \rangle$, even for method 4, which is why the $\chi^{2}_{\rm r}$ is still significantly larger than 1. The causes of these persistent biases are currently unknown, and require further investigation. They may be affected by the cross-correlation estimator; a move to the superior \citet{Landy1993} estimator should be explored in future work. Alternatively, the definition of target sample tomographic bins may be sub-optimal; wider equi-populated bins at higher redshifts will offer lower signal-to-noise of clustering redshift measurements, since those galaxies span a larger volume in a flux-limited survey where the number density falls with increasing redshift. Optimisation of the target tomography is another promising avenue for future work.
\begin{figure*}
	\centering
	\includegraphics[width=17cm]{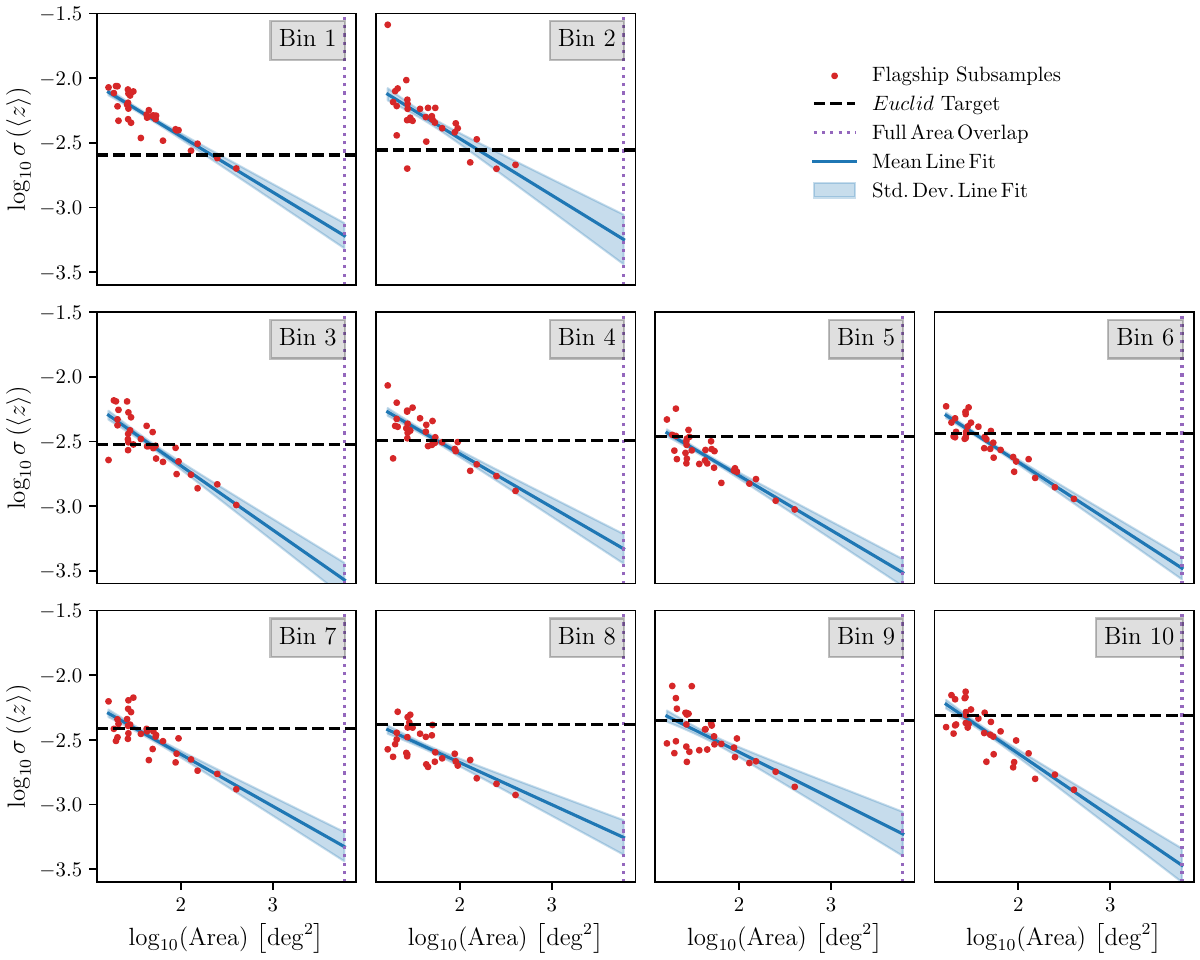}
	\caption{Uncertainty on the mean redshift for each \Euclid tomographic redshift bin shown as a function of tracer--target overlap area. The red points show the uncertainty on the mean redshift for each subregion, determined by fitting the shifted-true model (Sect. \ref{shiftedtruemodel}) to measured clustering redshifts (bias correction method 2, Sect. \ref{sec:cc_theory}), with the largest area points representing constraints from the full $402\,\rm{deg}^{2}$ Flagship footprint. These points are used to fit a power-law relationship between overlap area and mean redshift uncertainty, shown in blue (the line represents the mean fit and the envelope the uncertainty). Extrapolated to the full area overlap of BOSS, DESI, and \Euclid (the dotted purple line), the projected uncertainties on the mean are shown to be much smaller than the required uncertainties for \Euclid, itself indicated by the dashed black line.}
	\label{fig:nz_area_rescaling}
\end{figure*}

\subsection{Area rescaling}
\label{sec:area_rescaling_results}

A determination of whether or not clustering redshifts can constrain the $\sigma(\langle z\rangle)$ of \Euclid tomographic redshift bins to less than $0.002\,(1+z)$ requires extrapolation. We estimated the relation between $\sigma(\langle z\rangle)$ and tracer--target overlap area by measuring $\langle z\rangle$ within subregions of the Flagship region (see Fig. \ref{fig:bootstraps_area}). We fitted the $\sigma(\langle z\rangle)$ data with power laws and then extrapolated to the total projected overlap. The relations can be seen in Fig. \ref{fig:nz_area_rescaling} for the shifted-true model, calibrated by method 2 (Sect. \ref{sec:cc_theory}) clustering redshifts measured on the `mean Rubin' photometric catalogue. The extrapolation projects the uncertainties to be significantly smaller than the required uncertainties for \Euclid. We can be confident of this statement, as even the $402\,{\rm deg}^{2}$ Flagship area yields uncertainties approaching or surpassing the \Euclid requirement. Similar results are seen for the other three photometric catalogues. A possible cause for concern is the underestimation of the errors in the four highest-redshift bins, shown in Fig.~\ref{fig:cc_redshift_method}. This underestimation can be calibrated with comparisons to simulations (such as Flagship), and accommodated by widening the error bars. We find that these uncertainties are underestimated by a factor of about two. Since Fig.~\ref{fig:nz_area_rescaling} shows the projected errors for the full overlap region to be around an order of magnitude smaller than the \Euclid's required uncertainties, 
we are confident that these underestimated errors can be accommodated, and that the \Euclid target remains achievable.

\section{Conclusion}
\label{sec:conclusions}

We measured clustering redshifts in the Flagship simulation to test their uncertainties in determining the mean redshifts, $\langle z\rangle$, for \Euclid tomographic bins. The method uses cross-correlations with mock spectroscopic samples, modelled after BOSS, DESI, and \Euclid NISP-S. Clustering redshifts were determined using the \texttt{YAW} software, for transverse pair separations between $100\,{\rm kpc}$ and $1000\,{\rm kpc}$, using simple galaxy bias correction schemes. Simulated photometric samples were constructed using the DNF photometric redshift code \citep{Pocino2021}. The redshift determinations were constructed from two sets of training samples, one that is fully representative in redshift and magnitude and a second that has a completeness drop off in $I_{\scriptscriptstyle\rm E}$ magnitudes similar to surveys such as Rubin LSST (denoted with `Rubin').

The clustering redshift distributions were fitted with two models: The first modifies the true $n(z)$ with an amplitude and a shift parameter (the `shifted-true' model). The second fits a `suppressed-GP' model, taking advantage of the non-parametric fitting ability of GPs and suppressing low signal-to-noise and negative fluctuations with a S/N-dependent suppression function.

These two models were fitted to Flagship measurements over an area of $402\,{\rm deg}^{2}$. By making measurements of the clustering redshifts on subregions of the Flagship footprint, we established power-law relations between the uncertainty on the mean redshift and the spectroscopic--photometric overlapping area. We used these relations to extrapolate the uncertainty to the full expected overlap area for BOSS, DESI, and \Euclid (approximately $6000\,{\rm deg}^{2}$) and showed that both models achieve uncertainties on the mean redshifts of less than $0.002\,(1+z)$ -- well within the required uncertainties for the \Euclid FoM on dark energy of $>400$.

However, systematic biases currently dominate the mean redshift determination and are independent of the redshift distribution model. Determining and mitigating these sources of systematic biases will be critical for the usage of cross-correlation redshifts by \Euclid. The most difficult step for clustering redshift calibration is determining the galaxy bias of photometric samples. This is the primary source of systematic error and was shown to be the cause of systematic biases in the low-redshift bins. However, these biases persist in the high-redshift bins even when taking advantage of simulation information to correct exactly for photometric galaxy bias, meaning some other source of error is responsible.

Further studies should seek to characterise and mitigate these additional sources of systematic error, by exploring different scales, biasing models, or cross-correlation estimators. Photo-$z$ are limited to $z<1.6$ to ensure that the true redshifts of galaxies rarely exceed $z>1.8$ in the tenth tomographic bin, since the analysis was limited to the redshift regime where spectroscopic tracers were defined (i.e. $z<1.8$). In a future analysis, quasar large-scale structure tracers from BOSS, eBOSS, and DESI should allow for a measurement of the $n(z)$ for high-redshift tomographic bins, though the sparsity of quasars is likely to pose challenges. As simulations of quasar samples are more difficult to implement, since they are highly biased tracers with more complex selection functions, we have not included them in this analysis. Future studies should consider creating such samples so that the biases and uncertainties on these higher-redshift bins can be better determined.

This study is based on idealised assumptions where target and tracer samples suffer no observational complications, such as the spatial incompleteness of spectroscopic samples, tracer--target density correlations with Galactic foregrounds, or systematic depth variations. As such, future studies, and \Euclid photo-$z$ mocks in particular, should attempt to model these sources of systematic error. Furthermore, the photometric catalogues used in this study assume supplementary photometry from a Rubin-like survey; in reality, only the southern sky will be supplemented with Rubin, and the northern sky will use photometry from an array of imaging surveys such as CFHT, Pan-STARRS, and J-PAS. This is likely to result in systematic differences in the quality of photo-$z$ based on positions on the sky. To study these effects, realistic \Euclid mock catalogues that attempt to simulate these systematic errors across the full sky are required.


\section*{Acknowledgements}

K.~Naidoo acknowledges support from the Science and Technology Facilities Council grant ST/N50449X and from the (Polish) National Science Centre grant \#2018/31/G/ST9/03388. H.~Johnston acknowledges support from the Delta ITP consortium, a program of the Netherlands Organisation for Scientific Research (NWO) that is funded by the Dutch Ministry of Education, Culture and Science (OCW). O.~Lahav acknowledges support from an STFC Consolidated Grant ST/R000476/1. K.~Naidoo, H.~Johnston, B.~Joachimi, and O.~Lahav were supported by the UK Space Agency. H.~Hildebrandt is supported by a Heisenberg grant of the Deutsche Forschungsgemeinschaft (Hi 1495/5-1) as well as an ERC Consolidator Grant (No. 770935). J.~L. van den Busch acknowledges support from the European Research Council under grant numbers 770935.
\AckEC

\bibliographystyle{aa} 
\bibliography{bibliography,EuclidCCextrarefss}

\end{document}